\documentstyle[aps,prl,twocolumn,epsfig,floats]{revtex}

\begin{document}
\draft
\wideabs{%
\title{Non-Hermitian Random Matrix Theory and Lattice QCD with
  Chemical Potential}
\author{H.~Markum and R.~Pullirsch}
\address{Institut f\"ur Kernphysik, Technische Universit\"at
          Wien, A-1040 Vienna, Austria}
\author{T.~Wettig}
\address{Institut f\"ur Theoretische Physik, Technische
          Universit\"at M\"unchen, D-85747 Garching, Germany}
\date{10 May 1999}
\maketitle
\begin{abstract}
  In quantum chromodynamics (QCD) at nonzero chemical potential, the
  eigenvalues of the Dirac operator are scattered in the complex
  plane.  Can the fluctuation properties of the Dirac spectrum be
  described by universal predictions of non-Hermitian random matrix
  theory?  We introduce an unfolding procedure for complex eigenvalues
  and apply it to data from lattice QCD at finite chemical potential
  $\mu$ to construct the nearest-neighbor spacing distribution of
  adjacent eigenvalues in the complex plane.  For intermediate values
  of $\mu$, we find agreement with predictions of the Ginibre ensemble
  of random matrix theory, both in the confinement and in the
  deconfinement phase.
\end{abstract}
}

\narrowtext

Physical systems which are described by non-Hermi\-tian operators have
recently attracted a lot of attention.  They are of interest, e.g., in
dissipative quantum chaos \cite{Grob88}, disordered systems
\cite{Hata96,Efet97a}, neural networks \cite{Somm88}, and quantum
chromodynamics (QCD) at finite chemical potential \cite{Step96}.
Often they display unusual and unexpected behavior, such as a
delocalization transition in one dimension \cite{Hata96}.
Consequently, analytical efforts have been made to develop
mathematical methods to deal with non-Hermitian matrices, see, e.g.,
Refs.~\cite{Gini65,Fyod96,Efet97b,Fein97,Jani97}.

Of particular interest in the analysis of complicated quantum systems
are the properties of the eigenvalues of the Hamilton operator.  In
the Hermitian case, it has been shown for many different systems that
the spectral fluctuations on the scale of the mean level spacing are
given by universal predictions of random matrix theory (RMT)
\cite{Guhr98}.  In QCD, one studies the eigenvalues of the Dirac
operator, and it was demonstrated in lattice simulations that at zero
chemical potential, the local spectral fluctuation properties in the
bulk of the spectrum are reproduced by Hermitian RMT both in the
confinement and in the deconfinement phase \cite{Hala95}.

If one considers QCD at nonzero chemical potential, the Dirac operator
loses its Hermiticity properties so that its eigenvalues become
complex.  The aim of the present paper is to investigate whether
non-Hermitian RMT is able to describe the fluctuation properties of
the complex eigenvalues of the QCD Dirac operator.  The eigenvalues
are generated on the lattice for various values of $\mu$.  We define a
two-dimensional unfolding procedure to separate the average eigenvalue
density from the fluctuations and construct the nearest-neighbor
spacing distribution $P(s)$ of adjacent eigenvalues in the complex
plane.  The data are then compared to analytical predictions of
non-Hermitian RMT.

We start with a few definitions.  At $\mu\ne0$, the QCD Dirac operator
on the lattice in the staggered formulation is given by \cite{Hase83}
\begin{eqnarray}
  \label{Dirac}
  M_{x,y}(U,\mu) & = &
  \frac{1}{2a} \sum\limits_{\nu=\hat{x},\hat{y},\hat{z}}
  \left[U_{\nu}(x)\eta_{\nu}(x)\delta_{y,x\!+\!\nu}-{\rm H.c.}\right]
  \nonumber\\
  &&\hspace{-48pt}
  +\frac{1}{2a}\left[U_{\hat{t}}(x)\eta_{\hat{t}}(x)e^{\mu}
    \delta_{y,x\!+\!\hat{t}}
    -U_{\hat{t}}^{\dagger}(y)\eta_{\hat{t}}(y)
    e^{-\mu}\delta_{y,x\!-\!\hat{t}}\right]
\end{eqnarray}
with the link variables $U$ and the staggered phases $\eta$.  The
lattice constant is denoted by $a$, and color indices have been
suppressed.

We consider gauge group SU(3) which corresponds to the symmetry class
of the chiral unitary ensemble of RMT \cite{Verb94}.  At zero chemical
potential, all Dirac eigenvalues are purely imaginary, and the
nearest-neighbor spacing distribution $P(s)$ of the lattice data
agrees with the Wigner surmise of Hermitian RMT,
\begin{equation}
  \label{Wigner}
  P_{\rm W}(s)=\frac{32}{\pi^2}\,s^2\,e^{-4s^2/\pi}\:,
\end{equation}
both in the confinement and in the deconfinement phase \cite{Hala95}.
This finding implies strong correlations of the eigenvalues.  In
contrast, for uncorrelated eigenvalues $P(s)$ is given by the Poisson
distribution, $P_{\rm P}(s)=e^{-s}$.

For $\mu>0$, the eigenvalues of the matrix in Eq.~(\ref{Dirac}) move
into the complex plane.  If the real and imaginary parts of the
strongly correlated eigenvalues have approximately the same average
magnitude, the system should be described by the Ginibre ensemble of
non-Hermitian RMT \cite{Gini65}.  As is the case in Hermitian RMT, we
assume that the chiral structure of the problem does not affect the
spectral fluctuation properties in the spectrum bulk.

For a complex spectrum, we define $P(s)$ to represent the spacing
distribution of nearest neighbors in the complex plane, i.e., for each
eigenvalue $z_0$ one identifies the eigenvalue $z_1$ for which
$s=|z_1-z_0|$ is a minimum \cite{Grob88}.  Clearly, this prescription
is not unique, but it is the most natural choice and, more
importantly, the only choice for which analytical results are
available. After ensemble averaging, one obtains a function $P(s,z_0)$
which, in general, depends on $z_0$.  However, the dependence on $z_0$
can be eliminated by unfolding the spectrum, i.e., by applying a local
rescaling of the energy scale so that the average spectral density is
constant in a bounded region in the complex plane and zero outside.
This will be discussed below.  After unfolding, a spectral average
over $z_0$ yields $P(s)$.  Of course, the question of whether, for a
given data set, the average behavior of the spectral density can
indeed be separated from the fluctuations has to be answered
empirically.

In the Ginibre ensemble, the average spectral density is already
constant inside a circle and zero outside, respectively \cite{Gini65}.
In this case, unfolding is not necessary, and $P(s)$ is given by
\cite{Grob88}
\begin{equation}
  \label{Ginibre}
  P_{\rm G}(s)=c \, p(cs)
\end{equation}
with  
\begin{equation}
  p(s)=2s\lim_{N\to\infty}\left[\prod_{n=1}^{N-1}e_n(s^2)\,e^{-s^2}
  \right] \sum_{n=1}^{N-1}\frac{s^{2n}}{n!e_n(s^2)}\:,
\end{equation}
where $e_n(x)=\sum_{m=0}^n x^m/m!$ and $c=\int_0^\infty ds \, s \,
p(s)=1.1429...$.  This result holds for strongly non-Hermitian
matrices, i.e., for ${\rm Re}(z)\approx{\rm Im}(z)$ on average. In the
regime of weak non-Hermiticity \cite{Fyod96}, where the typical
magnitude of the imaginary parts of the eigenvalues is equal to the
mean spacing of the real parts, the RMT prediction deviates from
Eq.~(\ref{Ginibre}).  (Our definitions differ from those of
Ref.~\cite{Fyod96} by a factor of $i$ so that in our case it would be
more appropriate to speak of weak non-anti-Hermiticity.) We shall
comment on this regime below.  For uncorrelated eigenvalues in the
complex plane, the Poisson distribution becomes \cite{Grob88}
\begin{equation}
  \label{Poisson}
  P_{\bar{\rm P}}(s)=\frac{\pi}{2}\,s\,e^{-\pi s^2/4}\:.
\end{equation}

We now introduce a simple method to unfold complex spectra.  This
method will then be applied to the Dirac spectrum on the lattice at
various values of $\mu\neq 0$, and the resulting $P(s)$ will be
compared to Eqs.~(\ref{Ginibre}) and (\ref{Poisson}).  Assuming that
the spectral density has an average and a fluctuating part,
$\rho(x,y)=\rho_{\rm av}(x,y)+\rho_{\rm fl}(x,y)$, we need to find a
map
\begin{equation}
  \label{map}
  z'=x'+iy'=u(x,y)+iv(x,y)
\end{equation}
such that $\rho_{\rm av}(x',y')\equiv 1$.  Conservation of the
probability implies that $\rho_{\rm av}(x',y')dx'dy' = dx'dy' =
\rho_{\rm av}(x,y)dxdy$.  Hence, $\rho_{\rm av}(x,y)$ is the Jacobian
of the transformation from $(x,y)$ to $(x',y')$,
\begin{equation}
  \label{Jacobian}
  \rho_{\rm av}(x,y)=\left|\partial_xu\,\partial_yv-
    \partial_yu\,\partial_xv\right|\:.
\end{equation}
The left-hand side of Eq.~(\ref{Jacobian}) is given by the data.
There are infinitely many ways to choose the functions $u$ and $v$ in
Eq.~(\ref{map}) so that Eq.~(\ref{Jacobian}) is satisfied.  (In
general, a conformal map (\ref{map}) which fulfills
Eq.~(\ref{Jacobian}) can only be constructed if $\rho_{\rm av}(x,y)$
satisfies special conditions.)  We choose $y'=v(x,y)=y$ which yields
$\rho_{\rm av}(x,y)=|\partial_xu|$ and, thus,
\begin{equation}
  \label{x'}
  x'=u(x,y)=\int_{-\infty}^xdt\rho_{\rm av}(t,y)\:.
\end{equation}
This corresponds to a one-dimensional unfolding in strips parallel to
the real axis.  For a fixed bin in $y$, $\rho_{\rm av}(x,y)$ is
obtained by fitting $\rho(x,y)$ to a low-order polynomial.

Let us now discuss how the lattice data were obtained.  The
simulations were done with gauge group SU(3) on a $6^3\times4$ lattice
using $\beta=6/g^2=5.2$ in the confinement region and $\beta=5.4$ in
the deconfinement region for $N_f=3$ flavors of staggered fermions of
mass $ma=0.1$.  For each parameter set, we sampled 50 independent
configurations.  The gauge field configurations were generated at
$\mu=0$, and the chemical potential was added to the Dirac matrix
afterwards.  This procedure requires an explanation, since it is known
that the quenched approximation at $\mu\ne 0$ is unphysical
\cite{Step96}.  However, despite serious efforts \cite{Barb97} there
is currently no feasible solution to the problem of a complex weight
function in lattice simulations.  In a random matrix model, the statistical
effort to generate configurations including the complex Dirac determinant
was shown to grow exponentially with $\mu^2N$, where $N$ is the lattice
size \cite{Hala97}.

\begin{figure}[-b]
  \centerline{\epsfig{figure=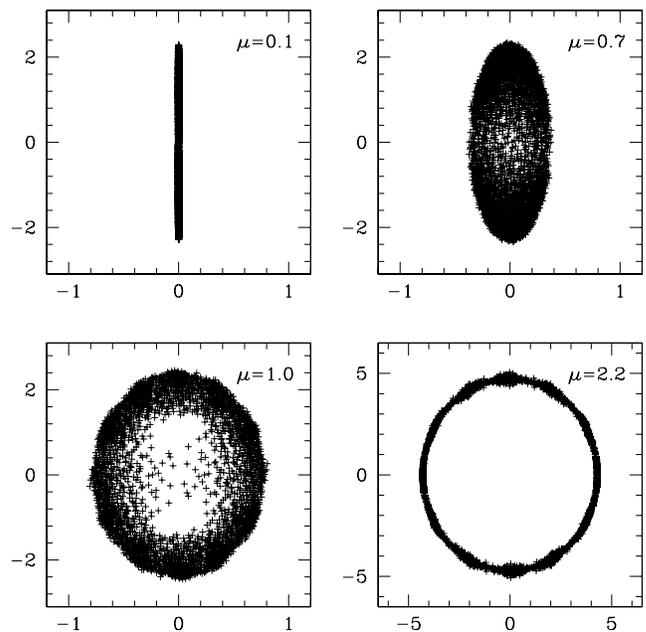,width=20pc}}
  \vspace*{5mm}
  \caption{Scatter plot of the eigenvalues of the Dirac operator (in
    units of $1/a$) in the complex plane at various values of $\mu$
    for a typical configuration (generated at $\mu=0$) in the
    confinement region at $\beta=5.2$.}
  \label{spectra}
\end{figure}

\begin{figure}[-b]
  \centerline{\epsfig{figure=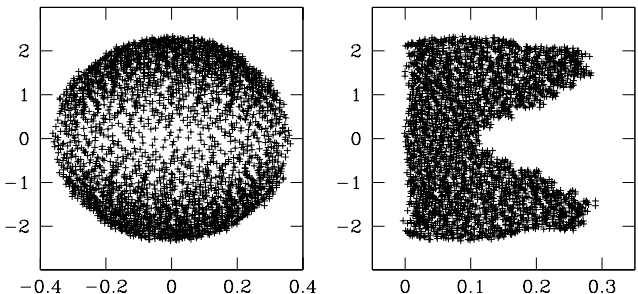,width=20pc}}
  \vspace*{5mm}
  \caption{Eigenvalue density of the Dirac matrix in the complex plane
    for a typical configuration at $\beta=5.2$ and $\mu=0.7$.  The
    left plot shows the eigenvalues before unfolding, the right plot
    after unfolding, respectively.}
  \label{unfspectra}
\end{figure}

Typical eigenvalue spectra are shown in Fig.~\ref{spectra} for four
different values of $\mu$ (in units of $1/a$) at $\beta=5.2$.  As
expected, the size of the real parts of the eigenvalues grows with
$\mu$, consistent with Ref.~\cite{Barb86}.  
Since the average spectral density is not constant, we have to
apply the unfolding method defined above.  Figure~\ref{unfspectra}
shows the effect of the unfolding procedure on the eigenvalue density
in the complex plane.
$P(s)$ is then constructed from the unfolded density and normalized
such that $\int_0^\infty ds\,s\,P(s)=1$.  

We have performed several checks of our unfolding method.  (i) If only
parts of the spectral support are considered, the results for $P(s)$
do not change.  This means that spectral ergodicity holds.  (ii) If
the spectral density has ``holes'' (see Fig.~\ref{spectra} for
$\mu=1.0$ and 2.2), we split the spectral support into several pieces
and unfold them separately.  This is justified by spectral ergodicity.
(iii) Unfolding each spectrum separately and ensemble unfolding yield
identical results for $P(s)$.  (iv) The results for $P(s)$ are stable
under variations of the degree of the fit polynomial and of the bin
sizes in $x$ and $y$.  Thus, we can conclude that it was indeed
possible with our unfolding method for complex spectra to separate the
average spectral density from the fluctuations, which is the
prerequisite for comparisons with analytical RMT \nolinebreak
predictions.

\begin{figure}
  \centerline{\epsfig{figure=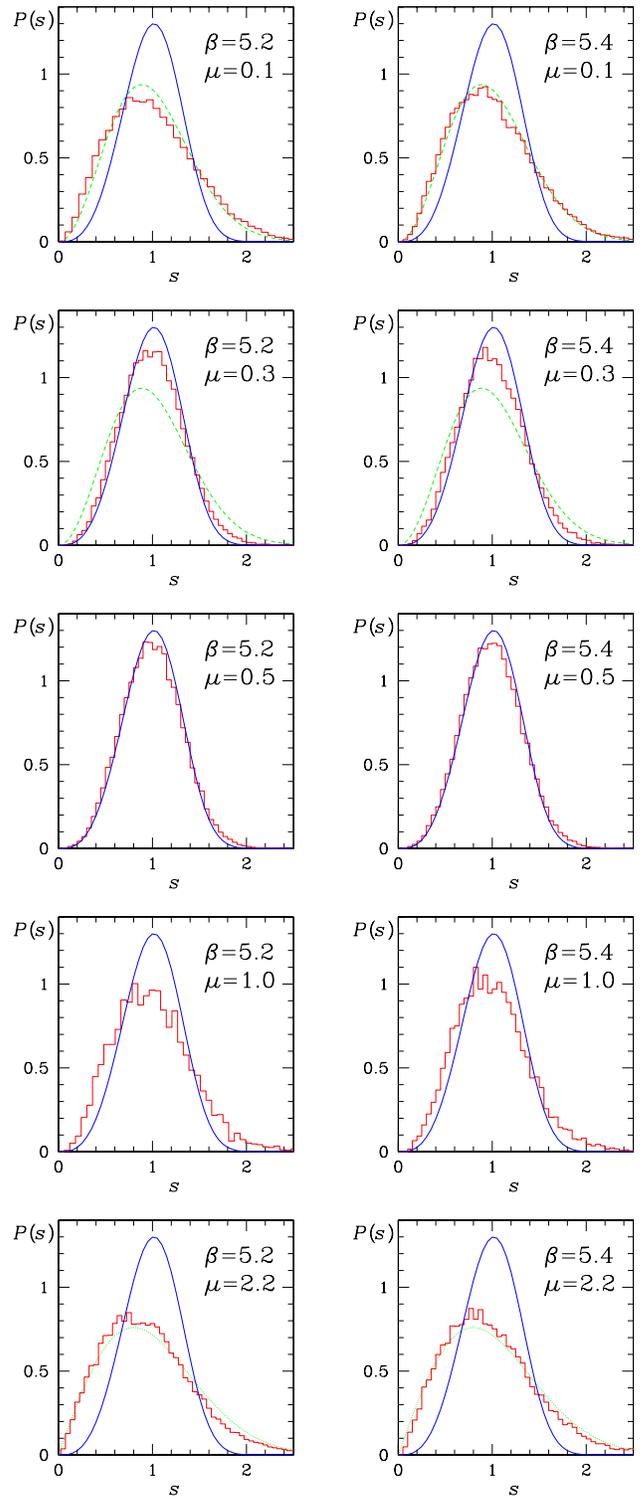}}
  \vspace*{5mm}
  \caption{Nearest-neighbor spacing distribution of the Dirac operator
    eigenvalues in the complex plane for various values of $\mu$ in
    the confinement ($\beta=5.2$, left) and deconfinement
    ($\beta=5.4$, right) phase. The histograms represent the lattice
    data. The solid curve is the Ginibre distribution of
    Eq.~(\protect\ref{Ginibre}), the short-dashed curve in the upper
    two rows the Wigner distribution of Eq.~(\protect\ref{Wigner}),
    and the dotted curve in the lowest row the Poisson distribution
    of Eq.~(\protect\ref{Poisson}), respectively.}
  \label{Psconf}
\end{figure}

Our results for $P(s)$ are shown in Fig.~\ref{Psconf}.  There are
minor quantitative but no qualitative differences between confinement
and deconfinement phases, which is consistent with our findings at
$\mu=0$ (second reference of \cite{Hala95}).  As a function of $\mu$,
we expect to find a transition from Wigner to Ginibre behavior in
$P(s)$.  This is indeed seen in the figures.  For $\mu=0.1$, the data
are still very close to the Wigner distribution of Eq.~(\ref{Wigner})
whereas for $0.5 \leq \mu \leq 0.7$ ($\mu=0.7$ not shown) we
observe nice agreement with the Ginibre distribution of
Eq.~(\ref{Ginibre}).  Values of $\mu$ in the crossover region between
Wigner and Ginibre behavior ($0.1<\mu<0.3$) correspond to the regime
of weak non-Hermiticity mentioned above.  In this regime, the
derivation of the spacing distribution is a very difficult problem,
and the only known analytical result is $P(s,z_0)$ for small $s$,
where $z_0$ is the location in the complex plane (i.e., no unfolding
is performed) \cite{Fyod96}.  The small-$s$ behavior of
Eqs.~(\ref{Wigner}) and (\ref{Ginibre}) is given by $P_{\rm
  W}(s)\propto s^2$ and $P_{\rm G} (s)\propto s^3$, respectively, and
in the regime of weak non-Hermiticity we have $P(s,z_0)\propto
s^\alpha$ (for $s\ll1$) with $2<\alpha<3$ \cite{Fyod96}.  This smooth
crossover from $\alpha=2$ to $\alpha=3$ is also observed in our
unfolded data.

For $\mu > 0.7$ the lattice results for $P(s)$ deviate substantially
from the Ginibre distribution.  It is also evident from the appearance
of the spectra for $\mu=1.0$ and 2.2 in Fig.~\ref{spectra} that the
global spectral density of the lattice data is very different from
that of the Ginibre ensemble.  While this does not immediately imply
that the local spectral fluctuations are also different, it is an
indication for qualitative changes.  In particular, the results for
$\mu=2.2$ in Fig.~\ref{Psconf} could be interpreted as Poisson
behavior, corresponding to uncorrelated eigenvalues. In the Hermitian
case at finite temperature, $T$, lattice simulations show a
transition to Poisson behavior only for $\beta\to\infty$ when the physical
box size shrinks and the theory becomes free \cite{Hala95}.  Here,
however, we have to remember that the chemical potential was neglected
in the generation of the equilibrium gauge fields so that conclusions
about the physical content of this observation should be drawn with
care. Specifically, we are not entitled to make definite statements on
possible connections between the finite-density phase transition
(expected at smaller values of $\mu$) and the deviations from Ginibre
behavior.  A more plausible explanation of the transition to Poisson
behavior is provided by the following two (related) observations.
First, for large $\mu$ the terms containing $e^\mu$ in
Eq.~(\ref{Dirac}) dominate the Dirac matrix, giving rise to
uncorrelated eigenvalues.  Second, for $\mu>1.0$ the fermion density
on the $6^3\times4$ lattice reaches its maximum value given by the
Pauli exclusion principle.  In any event, the mechanisms behind this
transition are interesting and deserve further study.

If a solution to the problem of the generation of configurations with
a complex weight function becomes available, it would be very
interesting to redo our analysis.  Since previous computations in the
deconfinement phase at $\mu = 0$ have verified the RMT predictions
\cite{Hala95}, and since the present simulations at $\beta = 5.4$ are
already in the deconfinement phase for all values of $\mu$, we expect
the observed Ginibre behavior to persist.  Of course, this should be
verified by a full finite-density simulation.

In conclusion, we have proposed a general unfolding procedure for the
spectra of non-Hermitian operators.  This procedure was applied to the
QCD lattice Dirac operator at finite chemical potential.  Agreement of
the nearest-neighbor spacing distribution with predictions of the
Ginibre ensemble of non-Hermitian RMT was found between $\mu=0.5$ and
$\mu=0.7$ in both confinement and deconfinement phases.  The deviations
from Ginibre behavior for smaller values of $\mu$ are well understood
whereas the deviations for larger values of $\mu$ toward a Poisson
distribution require a better understanding of QCD at finite density.
In this context, an interesting observation is that the results for
$P(s)$ in the non-Hermitian case are rather sensitive to $\mu$ whereas
they are very stable under variations of $T$ in the Hermitian case
\cite{Hala95}.  It would be interesting to apply our unfolding method
to other non-Hermitian systems \cite{Somm88,Soko88}.

\bigskip This work was supported by FWF project P10468-PHY and DFG
grant We 655/15-1.  We thank N. Kaiser, K. Rabitsch, and J.J.M.
Ver\-baar\-schot for discussions.

\end{document}